# A Survey of Machine Learning Algorithms for 6G Wireless Networks

**Anita Patil**
**Department of ECE, S G Balekundri Institute of Technology, Belagavi, KA, India- 590010.**
anitap@sgbit.edu.in

**Sridhar Iyer**
**Department of ECE, KLE Dr. M.S. Sheshgiri College of Engineering and Technology, Belagavi, KA, India- 590008.**
sridhariyer1983@klescet.ac.in

**Rahul Jashvantbhai Pandya**
**Indian Institute of Technology, Dharwad, WALMI Campus, PB Road, KA, India- 580011.**
rpandya@iitdh.ac.in

**Abstract:**

The primary focus of Artificial Intelligence/Machine Learning (AI/ML) integration within the wireless technology is to reduce capital expenditures, optimize network performance, and build new revenue streams. Replacing traditional algorithms with deep learning AI techniques have dramatically reduced the power consumption and improved the system performance. Further, implementation of ML algorithms also enables the wireless network service providers to (i) offer high automation levels from distributed AI/ML architectures applicable at the network edge, (ii) implement application-based traffic steering across the access networks, (iii) enable dynamic network slicing for addressing different scenarios with varying quality of service requirements, and (iv) enable ubiquitous connectivity across the various 6G communication platforms.

In this chapter, we review/survey the ML techniques which are applicable to the 6G wireless networks. and also list the open problems of research which require timely solutions.

**List of Abbreviations used in the chapter.**

| Abbreviation | Full Form |
|---|---|
| AMBC | Ambient Backscattering Communications |
| AI | Artificial Intelligence |
| APM | Amplitude Phase Modulation |
| CR | Cognitive Radio |
| DSA | Dynamic Spectrum Adaptation |
| DL | Deep Learning |
| DRL | Deep Reinforcement Learning |
| FL | Federated Learning |
| GAN | Generative Adversarial Networks |
| IM | Index Modulation |
| MDP | Markov Decision Process |
| ML | Machine Learning |
| NOMA | Non-Orthogonal Multiple Access |
| OFDMA | Orthogonal Frequency Division Multiple Access |
| QoS | Quality of Service |
| RKHS | Reproducing Kernel Hilbert Space |
| SM | Space Modulation |
| SSK | Space shift Keying |
| SWIPT | Simultaneous wireless information and Power Transfer |
| SR | Symbiotic Radio |
| SVM | Super Vector Machine |
| THz | Tera Hertz |

## 1. Overview of Machine learning in wireless Communication networks

With the exponential increase in the bandwidth demand and data traffic, there is an immediate requirement to serve this traffic through high-speed wireless communication networks. In turn, this requisites efficient software enabled intelligent algorithms, advanced physical layer solutions, and spectral bands at a higher frequency to fulfil the requirements of the next-generation users. The wireless communication research community has recently shown that the Tera-Hertz (THz) band is one of the promising bands to enable ultra-broadband wireless communication and minimize the spectrum scarcity issues (Zhao et. al., 2021).

The current wireless systems rely heavily on mathematical models; however, such models do not define the system structure accurately. Hence, the use of Machine learning (ML) techniques for wireless communication has gained momentum as these methods enable the attainment of the quality of service functionalities with advanced solutions (Ali et. al., 2020). Moreover, ML techniques provide the replacement of heuristic or Brute Force Algorithms for optimizing localized tasks and can also present adequate solutions that the existing mathematical model are unable to obtain. Currently, the ML algorithms are being deployed and trained statically at different management layers, core, radio base stations, and mobile devices. The dynamic deployment is envisioned to yield enhanced performance and utilization.

In general, the ML algorithms help in tasks such as, classification, regression, the interaction of an intelligent agent with the wireless environment (Syed et. al. 2019). In such operations, ML algorithms work in three different versions viz., supervised learning, unsupervised learning, and reinforcement learning. Few ML models such as, non-parametric Bayesian methods (Gaussian approach), are promising, especially in handling small, incrementally growing data sets; however, they have increased complexity compared to the parametric methods. Further, the Kernel Hilbert Space-based solutions have shown encouraging results in generating improved data rate, which is 10-100 times higher in comparison to the ones shown in the 5G wireless networks, simultaneously being computationally simple and scalable with lower approximation error. Federated Learning (FL) is an alternate distributed ML algorithm which enables mobile devices to collaboratively learn a shared ML model without data exchange among mobile devices (Marmol et. al., 2021). It is being analysed further to be considered as a next-generation solution for orientation, intrusion detection, mobility, and extreme event prediction. Reinforced Learning algorithms help in coding scheme selection, modulation, beam forming, and power control. In addition, physical layer optimization also exploits ML for multi-input and multi-output downlink beam forming. Implementing all the aforementioned ML algorithms at the end-user devices, needs the consideration of key parameters such as, cost, size, and power. Additional considerations in the simulation and the prototyping of ML at the end-user devices are to optimize the physical realization of the design and finding the inputs to the model (Dalal & Kushal, 2019).

The focus of this chapter is to bring out the importance of AI and ML in 6G wireless communication. ML is a component of AI although it endeavors to solve the problems based on historical or previous examples. Unlike AI applications, ML involves learning of hidden patterns within the data (data mining) and subsequently using the patterns to classify or predict an event related to the problem. Simply put, intelligent machines depend on the knowledge to sustain their functionalities and ML offers the same. In essence, ML algorithms are embedded into machines and data streams provided so that knowledge and information are extracted and fed into the system for faster and efficient management of processes (Ali et. al., 2020).

Demand for radio spectrum is increasing as the data traffic is increasing, and hence, massive connections with high quality of service have to be provided. Recent advances in ML have shown that ML will play a major role in solving multiple issues in wireless communication networks. To mention few, ML will provide ease in all sort of applications which were not enabled in the earlier generations such as, Augmented Reality (AR), Virtual Reality (VR), holographic telepresence, eHealth, wellness applications, Massive Robotics, Pervasive connectivity in smart Environment, etc. It is envisioned that ML will enable real time analysis and zero-touch operation, and will provide control in in the 6G networks (Zhao et. al., 2020). Mobile devices can assist and report to the network regarding the ML actions and predictions to aid efficient resource management. In order to manage the connection density, dynamic spectrum management has been proposed in the literature. The key enabling techniques for dynamic spectrum are i) Cognitive Radio ii) Symbiotic Radio, and iii) Blockchain Technology (Hong et. al., 2014; Hewa et. al. 2020).

The scarcity of available spectrum and underutilization of the allocated spectrum necessitates efficient techniques to manage the spectrum dynamically. In dynamic spectrum management, the concept of primary and secondary users exists wherein; secondary users do not have the authority to access the spectrum; however, they can access it whenever the primary spectrum is idle, and it can even be shared with the protection of primary users' service. This process enables the secondary users to transmit their data without the licence of spectrum.

In order to achieve dynamic spectrum allocation many algorithms are proposed in the literature. These algorithms not only address the issues of spectrum allocation but also issues such as, data security, optimization, power-efficiency, cost-efficiency, etc. Following are the ML algorithms which help in addressing all the aforementioned issues.

i. Supervised Learning

ii. Unsupervised learning

iii. Reinforced Learning

iv. Federated Learning

v. Kernel Hilbert Space

vi. Block Chain Technology

vii. Cognitive Radio

viii. Symbiotic Radio.

ix. THz Technology

x. Free Duplex

xi. Index Modulation

1. **Supervised Learning and Unsupervised Learning Algorithms**

The ML algorithms can be mainly classified as supervised or unsupervised. These two classes mainly differ in the labels of training data sets. In supervised ML output and input attributes are predetermined (Amanpreet et. al. 2016). The algorithms perform prediction and classification of the predetermined attributes, and their accuracy. For instance, if the input variable is $X$ and output variable is $Y$ then, the mapping function using by supervised learning will be $Y = F(X)$. The learning process stops when the algorithm achieves an acceptable level of performance. As detailed by (Chowdhury et. al., 2020), first analytical tasks are performed by supervised algorithms using the training data, and subsequently contingent functions are constructed for mapping variations of the attribute. These algorithms need pre-specifications of maximum settings to obtain the desired outcome and performance. With this approach in ML, it has been observed that the training subset of approximately 66% is rationale, and without demanding more computational time, outcome will be achieved. Further, the supervised learning algorithms can be further classified into the classification and regression algorithms.

- **Classification**: If output variable is a category such as, "red" or "blue" or "disease" and "no disease" the classification problem can be used (Amanpreet et. al. 2016).

- **Regression**: If the output value is a real value such as, "rupees" then, the regression problem can be used. Few popular examples of supervised ML algorithms are:

  ➢ Linear regression for regression problems.
  ➢ Random forest for classification and regression problems.
  ➢ Support vector machines (SVM) for classification problems.
  ➢ For resource allocation, and coding scheme selection in wireless communication, supervised learning algorithms are utilized.

**2. Semi-Supervised Machine Learning**

In these algorithms, large amount of input data $X$ and small amount of output data $Y$ are labelled. These problems lie in between the supervised and the unsupervised learning (Sheena & Sachin, 2019; Yogesh et. al., 2020).

**3. Unsupervised Learning**

In contrast to the supervised learning, unsupervised data learning comprises of pattern recognition without having a target attribute. All the variables used in the analysis are used as inputs, and the techniques are suitable for clustering and association mining techniques. According to (Amruthnath at. al., 2018), unsupervised learning algorithms are used to create labels in the data that are subsequently used to implement the supervised learning tasks. Clustering algorithms identify inherent groupings within the unlabeled data, and subsequently assign label to each data value. Further, unsupervised association mining algorithms identify rules that accurately represent the relationships between the attributes. Overall, the aim of unsupervised learning is to model the underlying structure or distribution in the data so as to learn more about the data. These algorithms do provide correct answers and there is no teacher; hence, they are called as 'Unsupervised Algorithms'. These algorithms can also be grouped into **Clustering** and **Association** Algorithms (Zhao et. al., 2021).

- **Clustering**: To discover the inherent groupings in the data. An example is based on purchasing behaviour grouping of customers.
- **Association**: To discover rules that describe large portions of the given data such as, people who buy X will also buy Y.

Few additional examples of the unsupervised learning Algorithms are:
  ➢ K-means for clustering problems.
  ➢ A priori algorithm for association rule learning problems.

- The commonly used techniques are: clustering, auto encoders (Kien et. al., 2020), deep belief nets, generative adversarial networks, and the expectation–maximization algorithm. It is also used in the physical layer for optimal modulation, channel-aware feature extraction, anomaly detection, localization etc.

**4. Reinforcement Algorithm**

Reinforcement learning (RL) is one of the basic ML paradigms along with supervised and unsupervised learning. It is concerned with the manner in which intelligent devices are required to make decisions or take actions in an environment in order to maximize the notion of cumulative reward. It differs from supervised learning as it does not need labelled input/output pairs to be presented, and does not need sub-optimal actions to be explicitly corrected. Instead, the focus is on finding a balance between exploration and exploitation. Markov Decision Process (MDP) is typically used to state the environment, as reinforcement learning algorithms make use of dynamic programming techniques. Reinforcement learning algorithms do not assume knowledge of an exact mathematical model of the MDP; however, they target large MDPs where exact mathematical models become infeasible (Vasileios. Et. al., 2021).

RL finds an application in many disciplines such as, game theory, control theory, simulation based optimization, multi-agent systems, swarm intelligence and wireless communication. For example, in the operational research and control literature, reinforcement learning is referred as *approximate dynamic programming,* or *neuro-dynamic programming.* The theory of optimal control includes reinforcement learning problems which is concerned with the characterization and existence of optimal solutions and algorithms for their exact computation. The problems are less concerned with approximation or learning, specifically in the absence of a mathematical model of the environment. Function approximation and use of samples are the powerful elements of reinforcement learning use to deal with large environments and optimization of performance. Reinforcement learning problem consists of an agent to interact with an environment and learns how to take actions. At each step of the learning process, state of the environment is observed by the agent which then takes action from the available set of actions, receives a numeric reward, and moves to the next state. Hence, the aim of the agent is to maximize long term cumulative reward. Problem such as resource allocation in wireless domain can be formulated as a reinforcement learning problem wherein, neural networks can be used as function approximators to learn values of each state or the rewards that are generated by the environment. Many problems in wireless networks such as, power control, beamforming, and modulation and coding scheme selection are solved by various deep reinforcement learning architectures. However, as a drawback, reliance on training is a major limitation of reinforcement learning. In this regard, recently, there has been advances towards reducing this reliance, specifically when dealing with extreme network situations. In particular, the concept of experienced deep reinforcement learning was proposed in which reinforcement learning is trained using Generative Adversarial Networks (GAN) that generate synthetic data to complement a limited, existing real dataset (Iqbal, 2021).

Two elements make RL powerful viz., the use of samples to optimize the performance and the use of function approximation to deal with large environments. RL can be used in following situations:

i. Only a simulation model of the environment is known i.e., subject of simulation optimization.

ii. A model of the environment is known and analytic solution is not available

iii. In 6G, RL can be used for resource allocation in cognitive network, beam forming, coding scheme selection, power control, channel modulation, etc.

RL Algorithms can also be categorised as described in the following sections.

## 5. Associative Reinforcement Learning

The learning system interacts in a closed loop with its environment and combines the facets of stochastic learning automata tasks.

## 6. Deep Reinforcement Learning

RL is extended in this approach by using a deep neural network without explicitly designing the state space. Learning ATARI games by Google Deep Mind has created increased attention to Deep Reinforcement Learning (DRL). It influences Markov decision models for selecting next 'action' based on the state transition models. In DRL, instead of mapping every solution, states are approximated or estimated by a neural network. In 6G wireless communications, efficient solutions for radio resource allocation are required and this is challenging as 6G wireless network will aim to serve wider variety of users in future radio resources which will be in extreme scarcity. As a solution, DRL can be implemented to obtain efficient solutions for the radio resource problems (Zhao et. al., 2020).

## 7. Inverse Reinforcement Learning

Purpose of inverse reinforcement learning is to imitate observed behaviour which is often optimal solution or close to optimal. Inverse reinforcement learning approach will have no reward function but a reward function will be inferred as an observed behaviour from expert.

## 8. Safe Reinforcement Learning

Safe RL is the process of learning policies that maximize the expectation of the return in problems wherein, it is important to ensure reasonable system performance, safety constraints during learning and deployment processes.

## 9. Federated Learning:

Federated learning (FL) is helpful is cases when it is difficult to assign the ML models to each mobile device and data center because in traditional centralized ML algorithms, it is required for each mobile device to transmit its collected data to the data center for training purpose. It is impractical for wireless mobile devices to transmit their local data for training ML models due to privacy issues. To overcome this traditional ML problem, FL, which is a distributed ML algorithm that enables devices to learn a shared ML model without data exchange among mobile devices, is implemented (Li et. al., 2021). In this approach, each mobile device and the data center will have their own ML models, and these ML models are referred to as 'local Model' and 'Global Model' for mobile device and data center, respectively.

Further, the training model of ML in FL approach is as follows:

a. Each mobile device uses the collected data to train the local FL model and sends the trained local FL model to the data center.
b. The data center integrates the local FL models to generate the global FL model and broadcasts it back to all mobile devices.

As it can be inferred from the FL process, the mobile devices are required to transmit the training parameters over the wireless links. The limited wireless bandwidth, imperfect transmission and dynamic wireless channels will affect the FL performance in a significant manner. However, many studies have shown that implementation of FL process by optimizing the wireless network. The main objective of FL is to protect the data owner's privacy. It aims to train a ML model with training data kept distributed at the clients in order to protect the data owners' privacy. The working idea illustration includes devices of the users are used to train local models and then the trained local models are sent to the base station for aggregation. In this process, the privacy of the user data is well preserved as the data are still maintained in the devices. In 6G networks, the architecture will be distributed wherein, the FL technology of AI from centralized cloud based model to decentralized devices will be necessitated. The AI computing tasks can be distributed to multiple decentralized edge nodes from a central node, and hence, FL is one of the vital ML methods to enable the deployment of accurately generalized models across multiple devices (Du et. al., 2020).

## 10. Kernel Hilbert space

High interference, which is a result of massive connectivity, will be a major performance bottleneck in 6G. Massive connectivity will involve serving extensive range of devices of various manufacturing qualities and this will lead to introduction of impairments which develop due to diverse objects introduced by non-ideal hardware. The major objects may be I/Q imbalance, non-linear characteristics, high-mobility especially in the industries where fixed solution may not be applicable, etc. To accomplish the promise of improvement in the data rate of 10-100 times in 6G as compared to the scenarios of 5G, the Reproducing kernel Hilbert space (RKHS) based solutions are predominantly useful owing to their computational simplicity (Salh et. al., 2021). RKHS methods have significantly lower approximation error and are scalable. 6G will potentially encounter the high interference non-Gaussian environments; however, RKHS based solutions will provide efficiency with lower approximation error. RKHS based approaches have recently appeared as a solution for many impairments in the framework of numerous applications in the next-generation wireless communication systems. Hence, the major issues in 6G such as, detection, tracking and localization will be addressed by various RKHS based solutions. In recent years, technological advances have ensured that DL method offers tremendous solutions, and it is largely used in wireless communications problems. Simultaneously, further improvement in the performance of RKHS based approaches could be achieved by the feature extraction methods using Monte-Carlo sampling. The extracted features can be utilized as input to the DL based slants to boost the act of models used in 6G. As compared to the classical DL algorithms, RKHS based DL algorithms ensure enhanced performance as there is an intrinsic regularization and support with strong analytical framework (Guo, 2021).

## 11. Cognitive Radio

The most widely accepted key technology to empower dynamic spectrum allocation is cognitive radio. Based on the interaction with surrounding environment, and also based on the awareness of its internal states such as, hardware and software architectures, user needs, spectrum use policy, cognitive radio can autonomously and dynamically adapt the system transmission strategies, bandwidth, transmit power, antenna beam, carrier frequency and modulation scheme. The transmission strategies are adjusted by computer software in the Software Defined Radio. Cognitive radio is capable to observe, analyze the observed environment through sensing, and process information through learning. It decides the transmission strategy through reasoning. Recent advances in research have shown that cognitive radio technology can be explored more for its inherent potential towards AI, and also its capacity to facilitate Dynamic Spectrum Allocation (DSA) very efficiently (Hong et. al., 2014).

Although most of the existing cognitive radio researchers till date have been focusing on the exploration and realization of cognitive capability to facilitate the DSA, recent research has been conducted to explore additional potential which is inherent in the cognitive radio technology through AI. A typical cognitive cycle for cognitive radio comprises of a Secondary User having cognitive capability which is essential to consistently and periodically observe the environment, and it obtains the information regarding interference temperature, spectrum holes, etc. It also determines the best operational features to optimize its own performance subjected to protecting the primary users, and according to these operational parameters, system configurations are conducted. The traffic statistics and channel fading statistics of the radio environment can be analyzed by the information time. Due to this, the cognitive radio equipment is able to learn and perform better in future Dynamic Spectrum Adaptation (DSA). However, implementing DSA with cognitive radio involves efforts from various research communities such as, communications, signal processing, computer networking, information theory and ML. Lastly, combining DSA with cognitive radio and its realization also fundamentally hinge on the inclination of regulators to open the spectrum for unlicensed access. Providentially, over the past decades, it has been observed that universal determinations from regulatory bodies are eradicating regulatory barriers to expedite DSA (Mollah et. al., 2019).

## 12. Block Chain Technology

In addition to the cryptocurrency, with its prominent features, blockchain has many uses including, smart contracts, financial services and IoT. Besides, blockchain can fetch new openings to improve the competence, and to lessen charge in the dynamic spectrum management. Also, blockchain is utilized to manage and share the spectrum resources, due to which central authority may get eliminated. Blockchain is fundamentally a distributed and open archive, in which transactions are firmly recognized in blocks. In the existing block, an exclusive indicator determined by transactions in the preceding block is recorded. Tampering with any transaction deposited in a previous block can be detected efficiently. The transactions commenced by one node are broadcast to other nodes, and a consensus algorithm is opted to conclude which node is approved to authorize the new block by affixing it to the blockchain. With the distributed authentication and record mechanism, blockchain will be robust against single point of failures, verifiable and transparent. Depending on the level decentralization, blockchain can be classified into consortium blockchain, public blockchain and private blockchain. A public blockchain can be tested and retrieved by all nodes in the network; whereas, a private blockchain or a consortium blockchain maintained by the permissioned nodes only. A smart contract, reinforced by the blockchain technology, is a self-executable connection with its sections being converted to programming writings and deposited in a transaction. Blockchain has been explored to provide support to various applications of IoT. As a decentralization entity, blockchain can be useful in integrating the various IoT devices and strongly stock the massive data formed by them. Moreover, blockchain is also used to accomplish the mobile edge computing resources to preserve the IoT devices with limited computation capacity. Blockchain provides security to dynamic spectrum access enabled by the spectrum auctions. In opportunistic spectrum access provided by sensors, block chain is used to intermediate the spectrum sensing service (Abdualgalil et. al., 2020). Specifically, blockchain can be used to:

**i. Establish a self-organized Spectrum Market:** In the management of spectrum resources, a self-organized spectrum market is essential with enhanced efficiency, with the reduced cost when compared to depending on a centralized authority. Dynamic spectrum management, transactions and their verification and smart contract combination will be used to block chain based self-organized spectrum market. The blockchain security assures that users can make the transactions without trust in each other. Thus, high bar to achieve spectrum resource accessibility can be reduced.

**ii. Create a secured database**: The information such as, outcomes of spectrum auctions and access records, historical sensing results can also be stored in the blockchain.

Overall, block chain technologies have been assumed to convey new opportunities to the dynamic spectrum management, to ensure progress in the avenues of security, autonomy, decentralization and administration cost reduction. Moreover, investigation on challenges such as, deployment, energy consumption, and scheme of blockchain network over the conventional cognitive radio network should also be carried out.

## 13. Symbiotic Radio

Symbiotic radio (SR) is an innovative technique that proposes the rewards and overcomes shortcomings of ambient backscattering communications (AmBC) and cognitive radio (CR). In building 6G as a spectrum and energy efficient communication system, SR is considered as one of the promising approach. In addition, Intelligent Reflecting Surfaces (IRS tool) can further play a role in improving the concert of the transmission by enhancing the backscattering link signal (Mingzhe et. al., 2021). Additional technologies, related to spectrum, in 6G are as follows:

**Terahertz**: Terahertz (THz) is measured as one of the vital technologies for the 6G wireless communications. 5G is described under 100 GHz as the millimeter-wave bands, whereas, 100 GHz-

3 THz is considered as the THz band in 6G. The beyond 90 GHz band is purely used for scientific service which has not been fully discovered. Therefore, it is proposed to support the increased wireless network capacity. THz also empowers the ultra-low latency communication and ultra-high bandwidth paradigms, which caters the desires of several evolving applications such as, IoT and autonomous driving. THz is especially adequate for high bit-rate short-range communications since, path loss increases with increase in frequency which makes it challenging for long-range communications.

**Free Duplex**: 6G will eradicate the difference between Frequency Division Duplex and Time Division Duplex, and the frequency sharing will be grounded on the necessities, which is known as free duplex. Hence, the spectrum resource allocation in 6G will be more effective and efficient. With the Free duplex technology, transmission rate and throughput in 6G can be increased and transmission latency will be reduced (Zhao et. al., 2021).

**Index Modulation:** Index modulation (IM) can progress the transmission rate as it carries the source information bits through the classical amplitude-phase modulation (APM) signals. Therefore, it can potentially be used in 6G. Studies suggest that information bits can be communicated through the index of the antennas in MIMO systems. This technique is known as space shift keying (SSK), and if SSK is combined with classical linear modulation, it is called as amplitude-phase modulation (APM). For example, space modulation (SM) is proposed based on the same idea as in SSK. In SM technology, the base information bits are distributed into two parts: the index of the transmit antennas and the other parts for the APM. Consequently, SM can meaningfully increase the transmission rate by sending the extra information bits through the traditional APM transceiver's antenna index. On one side of the antennas, further resource units can also be indexed to transmit the additional information bits. These resource units include, channel state, sub-carriers, and time slots. Index Modulation, along with Orthogonal frequency division multiplexing access (OFDMA), will be the key technology to suggestively upsurge the quantity output for supporting more users to access the 6G network (Zhao et. al., 2020).

Overall, 6G is expected to be a composite network where a huge selection of smart devices is connected to the system and are mandatory to interconnect with others anytime, and the life period of the battery-charging components is also vital to fulfill the restraints of ultra-low power consumption. To extend the life duration of numerous devices in the network, simultaneous wireless information and power transfer (SWIPT) technology is anticipated. SWIPT empowers sensors to be charged using wireless power transfer; thereafter, battery-free devices can be maintained in 6G, dropping the network's power consumption considerably. Consequently, performance on the sum rate, throughput, and outage probability for non-orthogonal multiple access (NOMA) networks with SWIPT are derived.

## 2. Applications of 6G

In the case of 6G networks, a varied series of AI applications will develop into 'connected intelligence (Shrestha et. al., 2019), later smoothing every facet of our day-to-day life. For instance:

i. Innovative AI methods can be hired in autonomy to save manpower or for network management. Overall, 6G invigorates smart healthcare by providing high-precision medical treatment, real-time health monitoring, and reliable privacy protection.

ii. With the arrival of 6G, Industry 4.0 will be completely comprehended as smart manufacturing and it will achieve high precision manufacturing.

iii. Intelligent robots coupled by pervasive 6G network empower manufacturing structures to carry out multifaceted and hazardous tasks devoid of risking people's life.

iv. The smart home that furnishes with smart IoT devices will deliver a contented living atmosphere to people, and 6G permits the smart home to confirm the occupants' safety.

v. In terms of traffic and transportation, the sophisticated sensing and planning algorithms can be deployed for traffic optimization.

vi. Other applications such as, smart grid and unmanned aerial vehicle will also be enhanced with the aid of 6G.

**Explainable Artificial Intelligence:** An enormous number of applications such as, remote surgery and autonomous driving exist in 6G eon. As these applications are attentively connected to humans' life, a small error may invite miserable accidents. Hence, it is very essential to mark an AI explainable for building faith between individuals and machines. Currently, most AI tactics in MAC and PYH layers of 5G wireless networks are incomprehensible. AI applications such as, remote surgery and autonomous driving will be an extensive part of 6G, which needs explaining ability to qualify trust. AI decisions should be understood by human experts, and must be explainable so as to be considered as reliable. Prevailing approaches, including didactic statements, visualization with case studies, hypothesis testing can improve the explaining ability of DL (Guo, 2019; Abhishek & Neha, 2021).

## 3. Open Research Avenues

Finally, we list the various open problems of research in regard to ML in 6G wireless networks

a. Implementation of automation requires higher data rates, flexibility, link preservation, scheduling, on demand beam forming-deep neural network with reinforcement learning algorithms need to be analysed for usage and deployment constraints have to be studied. In such case, solution to a challenge 'how to use deep reinforcement learning for the automation of 6G wireless network' will be an observable one.

b. Ultra-reliable low latency and interference management are the major requirements in open data access. To support the emerging machine critical applications, the transmission time delay of E-to-E (Equipment to Equipment) is expected to be less than 1ms. Satisfying low latency and high reliability needs with open data access in business oriented mobile network is an open problem in 6G.

c. In Application and Platform dependent models a design of device is necessary which can sense the environment and its history so that resources can be allocated dynamically and spectrum selection should also happen dynamically to avoid the congestion. Hence the transfer of models to highly resource constrained platforms and designing a method to dynamically select application and platform based models is a research challenge.

d. Implementation of ML algorithms for enhancing, automating and managing the 6G network performance is another research challenge.